\documentstyle [aps,multicol,prl]{revtex}
\begin{document}                                                    
\draft                                                      

\begin{multicols}{2}
                                                            
\noindent {\bf Comment on ``Spin Dependent Hopping and Colossal
Negative Magnetoresistance
in Epitaxial ${\rm \bf Nd_{0.52}Sr_{0.48}MnO_{3}}$
 Films in Fields up to 50 T''}
                                                            
\vspace{0.5cm}

Recently Wagner et al. \cite {wagner}
 proposed a modification of Mott's original model
 to explain the magnetoresistance
scaling in the ferromagnetic and paramagnetic regimes of the perovskite
 ${\rm Nd_{0.52}Sr_{0.48}MnO_{3}}$ films (in fields up to 50 T).
These authors claimed that there
is a hopping barrier which depends on the misorientation between 
the spins of electrons at the initial and
the final states in an elementary 
process. They further claimed that using the model they can explain the
observed scaling behavior-- negative-magnetoresistivity
scaling proportional to
the Brillouin function $\cal{B}$ in the ferromagnetic 
state and to ${\cal{B}}^2$ in the paramagnetic  state.
In this comment we argue that the {\it modification needed for Mott's
original model is different from that proposed by Wagner et al.} 
and further show that our picture will successfully explain the
observed scaling in the two regimes.

 Firstly, within a polaronic picture where hopping takes place from a
polarized cloud, the {\it wave function of the carrier in two dimensions
corresponds to a super localized carrier
and not to just a localized carrier}.
This claim of super localization follows from the correspondence of the
energy equation to that of a simple harmonic oscillator.
The polaronic free energy expression
 (up to a constant) that needs to be minimized
in two dimensions is given by
\begin{equation}                                          
\frac {\pi ^2 \hbar ^2}{2 m  R_{\xi} ^2 } + \frac{ \pi R_{\xi}^2}{a^2}
[ \Delta I - T \Delta S] ,
\label{Fmin} 
\end{equation}                                          
where $R_{\xi}$  is the radius of the polaron,
 $a$ is the lattice constant,
$\Delta I$ is the change in interaction energy due to polaron
formation, $\Delta S$ is the change in entropy,
 and $m= \hbar ^2 /(2t a^2)$ 
with $t$ being the hopping integral.
At $T=T_C$ 
we get a ferromagnetic transition because
the polarons try to align in the same direction and coalesce so as to
minimize the kinetic energy and because
the area spanned by the polarons is of the order of the area
of the system.
Also from  Eq.\ (\ref{Fmin})
it follows that the wavefunction is of the form 
$\exp [(- \alpha R^2/R_{\xi}^2)]$.
Thus the exponential part of the hopping conductivity is of the form 
$\exp {[- 2\alpha R^2/R_{\xi}^2 -W_{ij}/(k_{B} T)]}$
which when minimized in the usual fashion
yields $ \exp {[-(T_0/T)^{1/2}]}$.

Next, we observe that since the Hund's coupling constant is much 
larger than the hopping term $t$, in an elementary hopping process
the {\it hopping probability gets modified by a multiplicative term
$(1+M^2/{M_S}^2)/2$} with $M/M_S$ being magnetization fraction.
 Furthermore, {\it the potential energy difference
$W_{ij}$ between the 
two (initial and final) hopping sites does not change with
magnetization}.
 Because of the
large value of the Hund's coupling, as now known 
 \cite {degennes}, the mobile electron
 will always align its spin parallel to the localized spin.
 The above multiplicative
factor results from the sum of the probabilities for the following four 
processes:
(i)hopping from a randomly or paramagnetically
 oriented (P) site to a
 ferromagnetically aligned (F) site 
[$0.5 \times (1-M/M_S) \times (M/M_S)$];
(ii) hopping from a F
 site  to a P
site
[$0.5 \times (M/M_S) \times (1 - M/M_S)$];
(iii) hopping from a F
 site  to a F
 site [$(M/M_S) \times (M/M_S)$];
and (iv) hopping from a P
 site  to a P
 site [$0.5 \times (1-M/M_S) \times (1-M/M_S)$].
The above mentioned multiplicative term has also been deduced by Appel
\cite {appel}. 
Thus the hopping conductivity finally takes the form
\begin{eqnarray*}                                          
\sigma = e^2 R^2 \nu_{\rm ph}
\frac{(1+M^2/{M_S}^2 )}{2}
 N(E_F) 
\exp {[-(T_0/T)^{1/2}]} .
\end{eqnarray*}                                          

Now assuming that in the paramagnetic regime
the value of $M/M_S$ is small
we get the magnetoresistance to be proportional to
$ M^2/{M_S}^2 $. 
 Next we assume that the contribution to the magnetization
from the polarons is much larger than that from individual 
 spins
(which would be reasonable because of the size of the polarons and the
fact that at $T = T_C$ the area occupied by the polarons is 
of the order of the area of the system). Thus we approximate 
$M/M_S$ by the Brillouin function
 ${\cal{B}}[  g \mu _{B} J  B / (k_B T) ]$.
 
Lastly, {\it in the ferromagnetic regime the transport is band like and
not of hopping type} as assumed in 
Ref.\ \onlinecite {wagner}.
 In the ferromagnetic
regime when the magnetization is sizeable, a decrease in resistivity
due to an increase
in magnetic field would be linear in the increase 
in magnetization.  
As the magnetic field
in the ferromagnetic regime increases,
although the increase in the size of the polaron 
is small (because Zeeman energy is smaller than interaction energy),
more number of polarons get aligned
in the direction of the magnetic field 
thereby increasing the size of the
conducting domain.  Since the increase
in magnetization is mainly due to the magnetization of non-aligned
 polarons it can be taken to be proportional to
 $\cal{B}$.

At higher temperatures larger polarons
 get aligned due to the weight
of the Boltzmann factor. Thus the total spin $J$
 of the polaron,
as obtained from the Brillouin function,
 decreases as the temperature is lowered below $T_C$.


\vspace{0.3cm}

\noindent Sudhakar Yarlagadda 

 Saha Institute of Nuclear Physics, Calcutta, India

\vspace{0.3cm}

\noindent Received {\today}   


\end{multicols}

\begin{references}
                                       

\bibitem{wagner}
P. Wagner et al., Phys. Rev. Lett. {\bf 81}, 3980 (1998).

\bibitem{degennes}
P.-G. de Gennes, Phys. Rev. {\bf 118}, 141 (1960).

\bibitem{appel}
J. Appel, Phys. Rev. {\bf 141}, 506 (1966).

\end{references}
\end{document}